\documentclass[10pt,twocolumn,conference]{IEEEtran}
\usepackage{times}
\usepackage{amsbsy}
\usepackage{latexsym}
\usepackage{amssymb}
\usepackage{amsmath}
\usepackage[ansinew]{inputenc}
\usepackage[dvips ]{graphicx}
\input epsf
\usepackage{epic,eepic}
\usepackage{url}

\title{Adaptive Randomized Distributed Space-Time Coding in Cooperative MIMO Relay Systems}
\author{{Tong Peng $\S$, Rodrigo C. de Lamare $\S$ and Anke Schmeink
$\clubsuit$}\\
$\S$ Communications Research Group, Department of Electronics, University of York, York YO10 5DD, UK\\
$\clubsuit$ UMIC Research Centre, RWTH Aachen University, D-52056
Aachen, Germany \\
Emails: tp525@ohm.york.ac.uk; rcdl500@ohm.york.ac.uk,
schmeink@umic.rwth-aachen.de}

\begin{document}
\maketitle
\begin{abstract}
An adaptive randomized distributed space-time coding (DSTC) scheme
and algorithms are proposed for two-hop cooperative MIMO networks.
Linear minimum mean square error (MMSE) receivers and an
amplify-and-forward (AF) cooperation strategy are considered. In the
proposed DSTC scheme, a randomized matrix obtained by a feedback
channel is employed to transform the space-time coded matrix at the
relay node. Linear MMSE expressions are devised to compute the
parameters of the adaptive randomized matrix and the linear receive
filter. A stochastic gradient algorithm is also developed to compute
the parameters of the adaptive randomized matrix with reduced
computational complexity. We also derive the upper bound of the
error probability of a cooperative MIMO system employing the
randomized space-time coding scheme first.  The simulation results
show that the proposed algorithms obtain significant performance
gains as compared to existing DSTC schemes.
\end{abstract}

\section{Introduction}
Multiple-input and multiple-output (MIMO) communication systems
employ multiple collocated antennas at both the source node and the
destination node in order to obtain the diversity gain and combat
multi-path fading in wireless links. The different methods of
space-time coding (STC) schemes, which can provide a higher
diversity gain and coding gain compared to an uncoded system, are
also utilized in MIMO wireless systems for different numbers of
antennas at the transmitter and different conditions of the channel.
Cooperative MIMO systems, which employ multiple relay nodes with
antennas between the source node and the destination node as a
distributed antenna array, apply distributed diversity gain and
provide copies of the transmitted signals to improve the reliability
of wireless communication systems \cite{Scaglione}. Among the links
between the relay nodes and the destination node, cooperation
strategies, such as Amplify-and-Forward (AF), Decode-and-Forward
(DF), and Compress-and-Forward (CF) \cite{J. N. Laneman2004} and
various distributed STC (DSTC) schemes in \cite{J. N. Laneman2003},
\cite{Yiu S.} and \cite{RC De Lamare} can be employed.

The utilization of a distributed STC (DSTC) at the relay node in a
cooperative network, providing more copies of the desired symbols at
the destination node, can offer the system diversity gains and
coding gains to combat the interference. The recent focus on the
DSTC technique lies in the delay-toleration code design and the
full-diversity schemes design with the minimum outage probability.
In \cite{Sarkiss}, the distributed delay-tolerant version of the
Golden code \cite{Belfiore} is proposed, which can provide
full-diversity gain with a full coding rate. An opportunistic DSTC
scheme with the minimum outage probability is designed for a DF
cooperative network and compared with the fixed DSTC schemes in
\cite{Yulong}. An adaptive distributed-Alamouti (D-Alamouti) STBC
design is proposed in \cite{Abouei} for the non-regenerative
dual-hop wireless system which achieves the minimum outage
probability. DSTC schemes for the AF protocol are discussed in
\cite{Maham}-\cite{Sheng}. In \cite{Maham}, the GABBA STC scheme is
extended to a distributed MIMO network with full-diversity and
full-rate, while an optimal algorithm for design of the DSTC scheme
to achieve the optimal diversity and multiplexing tradeoff is
derived in \cite{Sheng}.

The performance of cooperative networks using different strategies
has been widely discussed in the literature. In \cite{Duong}, an
exact pairwise error probability of the D-Alamouti STBC scheme is
derived according to the position of the relay node. In
\cite{Minchul}, a bit error rate (BER) analysis of the
distributed-Alamouti STBC scheme is proposed. The difference between
these two works lies in the different cooperative schemes
considered. A maximum likelihood (ML) detection algorithm for a MIMO
relay system with DF protocol is derived in \cite{Sharma} with its
performance analysis as well. The symbol error rate and diversity
order upper bound for the scalar fixed-gain AF cooperative protocol
are given in \cite{Youngpil}. The use of single-antenna relay nodes
and the DF cooperative protocol is the main difference in scenario
between \cite{Youngpil} and this work. An STC encoding process is
implemented at the source node in \cite{Liang}, which decreases the
output of the system and increases the computational complexity of
the decoding at the destination node. In \cite{Unger}, the BER upper
bound is given without a STC scheme at the relay node.

In this paper, we propose an adaptive randomized distributed
space-time coding scheme and algorithms for a two-hop cooperative
MIMO relaying system with the AF protocol and linear MMSE receivers.
We focus on how the randomized matrix affects the DSTC during the
encoding and how to optimize the parameters in the matrix. It is
shown that the use of a randomized matrix benefits the performance
of the system by lowering the upper bound compared to using
traditional STC schemes. Linear MMSE expressions are devised to
compute the parameters of the adaptive randomized matrix and the
linear receive filter. Then an adaptive optimization algorithm is
derived based on the MSE criterion, with the stochastic gradient
(SG) algorithm in order to reduce the computational complexity of
the optimization process. The updated randomized matrix is
transmitted to the relay node through a feedback channel that is
assumed in this work error free and delay free. The upper bound
pairwise error probability of the randomized-STC schemes (RSTC) in a
cooperative MIMO system which employs multi-antenna relay nodes with
the AF protocol is also analyzed.

The paper is organized as follows. Section II introduces a two-hop
cooperative MIMO system with multiple relays applying the AF
strategy and the randomized DSTC scheme. In Section III the proposed
MMSE expressions and the SG algorithm for the randomized matrix are
derived, and the analysis of the upper bound of pairwise error
probability using the randomized D-STC is shown in Section IV.
Section V focus on the results of the simulations and Section VI
leads to the conclusion.

\section{Cooperative MIMO System Model}

\begin{figure}
\begin{center}
\def\epsfsize#1#2{1\columnwidth}
\epsfbox{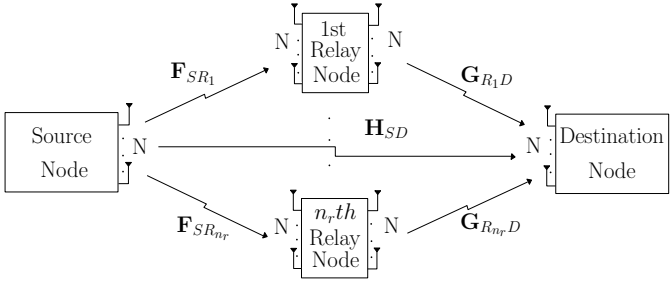}\vspace*{-0.5em} \caption{Cooperative MIMO System
Model with $n_r$ Relay nodes}\label{1}
\end{center}
\end{figure}

The communication system under consideration is a two-hop
cooperative MIMO system employing multiple relay nodes and
communicating over channels from the source node to the relay nodes
and the destination node, and from the relay nodes to the
destination nodes as shown in Fig. 1. A modulation scheme is used in
our system to generate the transmitted symbol vector ${\boldsymbol
s}[i]$ at the source node and all nodes have $N$ antennas. There are
$n_r$ relay nodes the employ the AF cooperative strategy as well as
a DSTC scheme. The system broadcasts symbols from the source to
$n_r$ relay nodes as well as to the destination node in the first
phase. The symbols are amplified and re-encoded at each relay node
prior to transmission to the destination node in the second phase.
We consider only one user at the source node in our system that has
$N$ Spatial Multiplexing (SM)-organized data symbols contained in
each packet. The received symbols at the $k-th$ relay node and the
destination node are denoted as ${\boldsymbol r}_{{SR}_{k}}$ and
${\boldsymbol r}_{SD}$, respectively, where $k=1,2,...,n_r$. The
received symbols ${\boldsymbol r}_{{SR}_{k}}$ are amplified before
mapped into an STC matrix. We assume that the synchronization at
each node is perfect. The received symbols at the destination node
and each relay node are described by
\begin{equation}\label{2.1}
{\boldsymbol r}_{{SR}_{k}}[i]  = {\boldsymbol F}_{k}[i]{\boldsymbol s}[i] + {\boldsymbol n}_{{SR}_{k}}[i],
\end{equation}
\begin{equation}\label{2.2}
{\boldsymbol r}_{SD}[i] = {\boldsymbol H}[i]{\boldsymbol s}[i] + {\boldsymbol n}_{SD}[i],
\end{equation}
\begin{equation*}
i = 1,2,~...~,N, ~~k = 1,2,~...~ n_{r},
\end{equation*}
where the $N \times 1$ vector ${\boldsymbol n}_{{SR}_{k}}[i]$ and
${\boldsymbol n}_{SD}[i]$ denote the zero mean complex circular
symmetric additive white Gaussian noise (AWGN) vector generated at
each relay and the destination node with variance $\sigma^{2}$. The
transmitted symbol vector ${\boldsymbol s}[i]$ contains $N$
parameters, ${\boldsymbol s}[i] = [s_{1}[i], s_{2}[i], ... ,
s_{N}[i]]$, which has a covariance matrix $E\big[ {\boldsymbol
s}[i]{\boldsymbol s}^{H}[i]\big] = \sigma_{s}^{2}{\boldsymbol I}$,
where $E[\cdot]$ stands for expected value, $(\cdot)^H$ denotes the
Hermitian operator, $\sigma_s^2$ is the signal power which we assume
to be equal to 1 and ${\boldsymbol I}$ is the identity matrix.
${\boldsymbol F}_k[i]$ and ${\boldsymbol H}[i]$ are the $N \times N$
channel gain matrices between the source node and the $k$th relay
node, and between the source node and the destination node,
respectively.

After processing and amplifying the received vector ${\boldsymbol
r}_{SR_k}[i]$ at the $k$th relay node, the signal vector
$\tilde{{\boldsymbol s}}_{SR_k}[i]={\boldsymbol
A}_{R_kD}[i]({\boldsymbol F}_{k}[i]{\boldsymbol s}[i] + {\boldsymbol
n}_{{SR}_{k}}[i])$ can be obtained and then forwarded to the
destination node. The amplified symbols in $\tilde{{\boldsymbol
s}}_{SR_k}[i]$ will be re-encoded by a $N \times T$ DSTC scheme
${\boldsymbol M}(\tilde{{\boldsymbol s}}[i])$ and then multiplied by
an $N \times N$ randomized matrix ${\boldsymbol {\mathfrak{R}}}[i]$
in \cite{Birsen Sirkeci-Mergen}, then forwarded to the destination
node. The relationship between the $k-th$ relay and the destination
node can be described as
\begin{equation}\label{2.3}
{\boldsymbol R}_{R_{k}D}[i] = {\boldsymbol G}_k[i]{\boldsymbol
{\mathfrak{R}}}[i]{\boldsymbol M}_{R_{k}D}[i] + {\boldsymbol
N}_{R_{k}D}[i],
\end{equation}
\begin{equation*}
    k = 1,2, ..., n_r,
\end{equation*}
where the $N \times T$ matrix ${\boldsymbol M}_{R_{k}D}[i]$ is the
DSTC matrix employed at the relay nodes whose elements are the
amplified symbols in $\tilde{{\boldsymbol s}}_{SR_k}[i]$. The $N
\times T$ received symbol matrix ${\boldsymbol R}_{R_{k}D}[i]$ in
(\ref{2.3}) can be written as an $NT \times 1$ vector ${\boldsymbol
r}_{R_{k}D}[i]$ given by
\begin{equation}\label{2.4}
{\boldsymbol r}_{R_{k}D}[i]  = {\boldsymbol
{\mathfrak{R}}}_{eq_k}[i]{\boldsymbol
G}_{{eq}_k}[i]\tilde{{\boldsymbol s}}_{SR_k}[i] + {\boldsymbol
n}_{R_{k}D}[i],
\end{equation}
where the block diagonal $NT \times NT$ matrix ${\boldsymbol
{\mathfrak{R}}}_{eq_k}[i]$ denotes the equivalent randomized matrix
and the $NT \times N$ matrix ${\boldsymbol G}_{eq_k}[i]$ stands for
the equivalent channel matrix which is the DSTC scheme ${\boldsymbol
M}(\tilde{{\boldsymbol s}}[i])$ combined with the channel matrix
${\boldsymbol G}_{R_{k}D}[i]$. The $NT \times 1$ equivalent noise
vector ${\boldsymbol n}_{R_{k}D}[i]$ generated at the destination
node contains the noise parameters in ${\boldsymbol N}_{R_{k}D}[i]$.
After rewriting ${\boldsymbol R}_{R_{k}D}[i]$ we can consider the
received symbol vector at the destination node as a $N(n_r+1)$
vector with two parts, one is from the source node and another one
is the superposition of the received vectors from each relay node.
Therefore, the received symbol vector for the cooperative MIMO
system can be written as
\begin{equation}\label{2.5}
\begin{aligned}
{\boldsymbol r}[i]  &=
\left[\begin{array}{c} {\boldsymbol H}[i]{\boldsymbol s}[i]  \\ \sum_{k=1}^{n_r}{\boldsymbol {\mathfrak{R}}}_{eq_k}[i]{\boldsymbol G}_{{eq}_k}[i]\tilde{{\boldsymbol s}}_{SR_k}[i] \end{array} \right] + \left[\begin{array}{c}{\boldsymbol n}_{SD}[i] \\ {\boldsymbol n}_{RD}[i] \end{array} \right] \\
& = {\boldsymbol D}_D[i]\tilde{\boldsymbol s}_D[i] + {\boldsymbol n}_D[i],
\end{aligned}
\end{equation}
where the $(T + 1)N \times (n_r + 1)N$ block diagonal matrix
${\boldsymbol D}_D[i]$ denotes the channel gain matrix of all the
links in the network which contains the $N \times N$ channel
coefficients matrix ${\boldsymbol H}[i]$ between the source node and
the destination node, the $NT \times N$ equivalent channel matrix
${\boldsymbol G}_{{eq}_k}[i]$ for $k=1,2,...,n_r$ between each relay
node and the destination node. The $(n_r + 1)N \times 1$ noise
vector ${\boldsymbol n}_D[i]$ contains the received noise vector at
the destination node and the amplified noise vectors from each relay
node, which can be modeled as additive white Gaussian noise (AWGN)
with zero mean and covariance matrix
$\sigma^{2}(1+\parallel{\boldsymbol
{\mathfrak{R}}}_{eq_k}[i]{\boldsymbol G}_{{eq}_k}[i]{\boldsymbol
A}_{R_kD}[i]\parallel^2_F){\boldsymbol I}$, where
$\parallel{\boldsymbol X}\parallel_F=\sqrt{{\rm Tr}({\boldsymbol
X}^H\cdot{\boldsymbol X})}=\sqrt{{\rm Tr}({\boldsymbol
X}\cdot{\boldsymbol X}^H)}$ is the Frobenius norm.

\section{Design of Linear MMSE Receivers and Randomized Matrices}

In this section, we design an adaptive linear MMSE receive filter
and an MMSE randomized matrix for use with the proposed DSTC scheme.
An adaptive SG algorithm \cite{S. Haykin} for determining the
parameters of the randomized matrix with reduced complexity is also
devised. The DSTC scheme used at the relay node employs an MMSE
randomized matrix, which is computed at the destination node and
obtained by a feedback channel and processes the data symbols prior
to transmission to the destination node.

\subsection{Optimization Method Based on the MSE Criterion}

Let us consider the MMSE design of the receive filter and the
randomized matrix according to the optimization problem
\begin{equation*}
    [{\boldsymbol W}[i],{{\boldsymbol {\mathfrak{R}}}_{eq}}[i]] = \arg\min_{{\boldsymbol W}[i], {{\boldsymbol {\mathfrak{R}}}_{eq}}[i]} E\left[\|{\boldsymbol s}[i]-{\boldsymbol W}^H[i]{\boldsymbol r}[i]\|^2\right],
\end{equation*}
where ${\boldsymbol r}[i]$ is the received symbol vector at the destination node which contains the randomized matrix to be optimized. If we only consider the received symbols from the relay node, the received symbol vector at the destination node can be derived as
\begin{equation}\label{4.1}
\begin{aligned}
    {\boldsymbol r}[i] &={\boldsymbol D}_D[i]\tilde{\boldsymbol s}_D[i] + {\boldsymbol n}_D[i]\\
    &={{\boldsymbol {\mathfrak{R}}}_{eq}}[i]{\boldsymbol G}_{eq}[i]{\boldsymbol A}[i]{\boldsymbol F}[i]{\boldsymbol s}[i]+{{\boldsymbol {\mathfrak{R}}}_{eq}}[i]{\boldsymbol G}_{eq}[i]{\boldsymbol A}[i]{\boldsymbol n}_{SR}[i]\\
    &~~~+{\boldsymbol n}_{RD}[i]\\
    &={{\boldsymbol {\mathfrak{R}}}_{eq}}[i]{\boldsymbol C}[i]{\boldsymbol s}[i]+{\boldsymbol n}_D[i],
\end{aligned}
\end{equation}
where ${\boldsymbol C}[i]$ is an $NT \times N$ matrix that contains
all the complex channel gains and the amplified matrix assigned to
the received vectors at the relay node, and the noise vector
${\boldsymbol n}_D$ is a Gaussian noise with zero mean and variance
$\sigma^{2}(1+\parallel{\boldsymbol
{\mathfrak{R}}}_{eq}[i]{\boldsymbol G}_{eq}[i]{\boldsymbol
A}[i]\parallel^2_F)$. We can then recast the optimization as
\begin{equation}\label{4.2}
\begin{aligned}
    &[{\boldsymbol W}[i],{{\boldsymbol {\mathfrak{R}}}_{eq}}[i]] =\\& \arg\min_{{\boldsymbol W}[i], {{\boldsymbol {\mathfrak{R}}}_{eq}}[i]} E\left[\|{\boldsymbol s}[i]-{\boldsymbol W}^H[i]({{\boldsymbol {\mathfrak{R}}}_{eq}}[i]{\boldsymbol C}[i]{\boldsymbol s}[i]+{\boldsymbol n}_D[i])\|^2\right].
\end{aligned}
\end{equation}
By expanding the righthand side of (\ref{4.2}) and taking the
gradient with respect to ${\boldsymbol W}^*[i]$ and equating the
terms to zero, we can obtain the linear MMSE receive filter
\begin{equation}\label{4.3}
    {\boldsymbol W}[i]=\left(E\left[{\boldsymbol r}[i]{\boldsymbol r}^H[i]\right]\right)^{-1}E\left[{\boldsymbol r}[i]{\boldsymbol s}^H[i]\right],
\end{equation}
where the first term denotes the inverse of the auto-correlation
matrix and the second one is the cross-correlation matrix. Define
$\tilde{\boldsymbol r}={\boldsymbol C}[i]{\boldsymbol
s}[i]+{\boldsymbol C}[i]{\boldsymbol n}_{SR}$, then the randomized
matrix can be calculated by taking the gradient with respect to
${\boldsymbol {\mathfrak{R}}}^*[i]$ and equating the terms to zero,
resulting in
\begin{equation}\label{4.4}
    {\boldsymbol {\mathfrak{R}}}[i]=\left({\boldsymbol W}^H[i](E\left[\tilde{\boldsymbol r}[i]\tilde{\boldsymbol r}^H[i]\right]){\boldsymbol W}[i]\right)^{-1}E\left[{\boldsymbol s}[i]\tilde{\boldsymbol r}^H[i]\right]{\boldsymbol W}[i],
\end{equation}
where $E\left[\tilde{\boldsymbol r}[i]\tilde{\boldsymbol
r}^H[i]\right]$ is the auto-correlation of the space-time coded
received symbol vector at the relay node, and $E\left[{\boldsymbol
s}[i]\tilde{\boldsymbol r}^H[i]\right]$ is the cross-correlation.
The expression above requires a matrix inversion with a high
computational complexity.

\subsection{Adaptive Randomized Matrix Optimization Algorithm}

In order to reduce the computational complexity and achieve the
optimal performance, an adaptive randomized matrix optimization
(ARMO) algorithm based on an SG algorithm is devised. The MMSE
problem is derived in (\ref{4.2}), and the MMSE filter matrix can be
calculated by (\ref{4.3}) first during the optimization process. The
simple ARMO algorithm can be obtained by taking the instantaneous
gradient term of (\ref{4.2}) with respect to the randomized matrix
${{\boldsymbol {\mathfrak{R}}}_{eq}}^*[i]$, which is given by
\begin{equation}\label{4.5}
\begin{aligned}
    & \nabla {\rm L}_{{{\boldsymbol {\mathfrak{R}}}_{eq}}^*[i]} \\ &= \nabla E\left[\|{\boldsymbol s}[i]-{\boldsymbol W}^H[i]({{\boldsymbol {\mathfrak{R}}}_{eq}}[i]{\boldsymbol C}[i]{\boldsymbol s}[i]+{\boldsymbol n}_D[i])\|^2\right]_{{{\boldsymbol {\mathfrak{R}}}_{eq}}^*[i]}\\
    &= -({\boldsymbol s}[i]-{\boldsymbol W}^H[i]{\boldsymbol r}[i]){\boldsymbol s}^H[i]{\boldsymbol C}^H[i]{\boldsymbol W}[i]\\
    &= -{\boldsymbol e}[i]{\boldsymbol s}^H[i]{\boldsymbol C}^H[i]{\boldsymbol W}[i],
\end{aligned}
\end{equation}
where ${\boldsymbol e}[i]$ stands for the detected error vector.
After we computing (\ref{4.5}), the ARMO algorithm can be obtained
by introducing a step size into an SG algorithm to update the result
until the convergence is reached as given by
\begin{equation}\label{4.7}
    {\boldsymbol {\mathfrak{R}}}[i+1]={\boldsymbol {\mathfrak{R}}}[i]+\mu({\boldsymbol e}[i]{\boldsymbol s}^H[i]{\boldsymbol C}^H[i]{\boldsymbol W}[i]),
\end{equation}
where $\mu$ stands for the step size of the ARMO algorithm. The
complexity of calculating the randomized matrix is ${\rm O}(2N)$,
which is much less than that of the calculation method derived in
(\ref{4.4}). As mentioned in Section I, the randomized matrix will
be sent back to the relay nodes via a feedback channel which is
assumed to be error-free in this work. However, in practical
circumstances, the errors caused by the broadcasting and the
diversification of the feedback channel with time changes will
affect the accuracy of the received randomized matrix at the relay
nodes.

\section{Probability of Error Analysis}

In this section, the upper bound of the pairwise error probability
of the system employing the randomized DSTC will be derived. As we
mentioned in the first section, the randomized matrix will be
considered in the derivation as it affects the performance by
reducing the upper bound of the pairwise error probability. For the
sake of simplicity, we consider a 2 by 2 MIMO system with 1 relay
node, and the direct link is ignored in order to concentrate on the
effect of the randomized matrix. The expression of the upper bound
is also stable for the increase of the system size and the number of
relay nodes.

Consider an $N \times N$ STC scheme we use at the relay node with
$L$ codewords. The codeword ${\boldsymbol C}^1$ is transmitted and
decoded to another codeword ${\boldsymbol C}^i$ at the destination
node, where $i=1,2,...,L$. According to \cite{Hamid}, the
probability of error can be upper bounded by the sum of all the
probabilities of incorrect decoding, which is given by
\begin{equation}\label{3.1}
    {\rm P_e} \leq \sum_{i=2}^L {\rm P}({\boldsymbol C}^1\rightarrow{\boldsymbol C}^i).
\end{equation}
Assuming the codeword ${\boldsymbol C}^2$ is decoded at the
destination node and we know the channel information perfectly at
the destination node, we can derive the pairwise error probability
as
\begin{equation}\label{3.2}
\begin{aligned}
    {\rm P}&({\boldsymbol C}^1\rightarrow{\boldsymbol C}^2\mid{{\boldsymbol {\mathfrak{R}}}})\\&={\rm P}(\parallel{\boldsymbol R}^1-{\boldsymbol G}{{\boldsymbol {\mathfrak{R}}}}{\boldsymbol C}^1\parallel^2_F-\parallel{\boldsymbol R}^1-{\boldsymbol G}{{\boldsymbol {\mathfrak{R}}}}{\boldsymbol C}^2\parallel^2_F>0\mid{{\boldsymbol {\mathfrak{R}}}_{eq}})\\&={\rm P}(\parallel{\boldsymbol r}^1-{{\boldsymbol {\mathfrak{R}}}_{eq}}{\boldsymbol G}_{eq}{\boldsymbol F}{\boldsymbol s}^1\parallel^2_F\\&~~~~~~~~~~~~~~~~-\parallel{\boldsymbol r}^1-{{\boldsymbol {\mathfrak{R}}}_{eq}}{\boldsymbol G}_{eq}{\boldsymbol F}{\boldsymbol s}^2\parallel^2_F>0\mid{{\boldsymbol {\mathfrak{R}}}_{eq}}),
\end{aligned}
\end{equation}
where ${\boldsymbol F}$ and ${\boldsymbol G}_{eq}$ stand for the
channel coefficient matrix between the source node and the relay
node, and between the relay node and the destination node,
respectively. The randomized matrix is denoted by ${{\boldsymbol
{\mathfrak{R}}}_{eq}}$. Define ${\boldsymbol H} = {\boldsymbol
G}_{eq}{\boldsymbol F}$, which stands for the total channel
coefficients matrix. After the calculation, we can transfer the
pairwise error probability expression in (\ref{3.2}) to
\begin{equation}\label{3.3}
    {\rm P}({\boldsymbol C}^1\rightarrow{\boldsymbol C}^2\mid{{\boldsymbol {\mathfrak{R}}}_{eq}})={\rm P}(\parallel{{\boldsymbol {\mathfrak{R}}}_{eq}}{\boldsymbol H}({\boldsymbol s}^1-{\boldsymbol s}^2)\parallel^2_F<{\boldsymbol Y}),
\end{equation}
where ${\boldsymbol Y}={\rm Tr}({\boldsymbol n}^{1^H}{{\boldsymbol
{\mathfrak{R}}}_{eq}}{\boldsymbol H}({\boldsymbol s}^1-{\boldsymbol
s}^2)+({{\boldsymbol {\mathfrak{R}}}_{eq}}{\boldsymbol
H}({\boldsymbol s}^1-{\boldsymbol s}^2))^H{\boldsymbol n}^1)$, and
${\boldsymbol n}^1$ denotes the noise vector at the destination node
with zero mean and covariance matrix
$\sigma^{2}(1+\parallel{\boldsymbol {\mathfrak{R}}}_{eq}{\boldsymbol
G}_{eq}\parallel^2_F){\boldsymbol I}$. By making use of the Q
function, we can derive the error probability function as
\begin{equation}\label{3.4}
    {\rm P}({\boldsymbol C}^1\rightarrow{\boldsymbol C}^2\mid{{\boldsymbol {\mathfrak{R}}}_{eq}})={\rm Q}\left(\sqrt{\frac{\gamma}{2}}\parallel{{\boldsymbol {\mathfrak{R}}}_{eq}}{\boldsymbol H}({\boldsymbol s}^1-{\boldsymbol s}^2)\parallel_F\right),
\end{equation}
where
\begin{equation}\label{3.4.2}
    {\rm Q}=\frac{1}{\sqrt{2\pi}}\int^\infty_x\exp\left(-\frac{u^2}{2}\right){\rm d}u,
\end{equation}
and $\gamma$ is the received SNR at the destination node assuming
the transmit power is equal to 1.

In order to obtain the upper bound of ${\rm P}({\boldsymbol
C}^1\rightarrow{\boldsymbol C}^2\mid{{\boldsymbol
{\mathfrak{R}}}_{eq}})$ we expand the formula
$\parallel{{\boldsymbol {\mathfrak{R}}}_{eq}}{\boldsymbol
H}({\boldsymbol s}^1-{\boldsymbol s}^2)\parallel^2_F$. Let
${\boldsymbol U}^H{\boldsymbol \Lambda}_s{\boldsymbol U}$ be the
eigenvalue decomposition of $({\boldsymbol s}^1-{\boldsymbol
s}^2)^H({\boldsymbol s}^1-{\boldsymbol s}^2)$, where ${\boldsymbol
U}$ is a Hermitian matrix and ${\boldsymbol \Lambda}_s$ contains all
the eigenvalues of the difference between two different codewords
${\boldsymbol s}^1$ and ${\boldsymbol s}^2$. Let ${\boldsymbol
V}^H{\boldsymbol \Lambda}_{\mathfrak{R}}{\boldsymbol V}$ stand for
the eigenvalue decomposition of $({\boldsymbol
{\mathfrak{R}}}_{eq}{\boldsymbol H}{\boldsymbol U})^H{\boldsymbol
{\mathfrak{R}}}_{eq}{\boldsymbol H}{\boldsymbol U}$, where
${\boldsymbol V}$ is a random Hermitian matrix and ${\boldsymbol
\Lambda}_{\mathfrak{R}}$ is the ordered diagonal eigenvalue matrix.
Therefore, the probability of error can be written as
\begin{equation}\label{3.5}
    {\rm P}({\boldsymbol C}^1\rightarrow{\boldsymbol C}^2\mid{{\boldsymbol {\mathfrak{R}}}_{eq}})={\rm Q}\left(\sqrt{\frac{\gamma}{2}\sum^{NT}_{m=1}\sum^N_{n=1}\lambda_{\mathfrak{R}_m}\lambda_{s_n}
    |\xi_{n,m}|^2}\right),
\end{equation}
where $\xi_{n,m}$ is the $(n,m)-th$ element in ${\boldsymbol V}$,
and $\lambda_{\mathfrak{R}_m}$ and $\lambda_{s_n}$ are eigenvalues
in ${\boldsymbol \Lambda}_{\mathfrak{R}}$ and ${\boldsymbol
\Lambda}_s$, respectively. According to \cite{Hamid}, a good upper
bound assumption of the Q function is given by
\begin{equation}\label{3.5.2}
    {\rm Q}(x)\leq\frac{1}{2}e^{\frac{-x^2}{2}}.
\end{equation}
Thus, we can derive the upper bound of pairwise error probability
for a randomized STC scheme as
\begin{equation}\label{3.6}
    {\rm P}({\boldsymbol C}^1\rightarrow{\boldsymbol C}^2\mid{{\boldsymbol {\mathfrak{R}}}_{eq}})\leq\frac{1}{2}\exp\left(-\frac{\gamma}{4}\sum^{NT}_{m=1}\sum^N_{n=1}\lambda_{\mathfrak{R}_m}
    \lambda_{s_n}|\xi_{n,m}|^2\right),
\end{equation}
while the upper bound of the error probability expression for a
traditional STC is given by
\begin{equation}\label{3.7}
    {\rm P}({\boldsymbol C}^1\rightarrow{\boldsymbol C}^2\mid{\boldsymbol H}_{eq})\leq\frac{1}{2}\exp\left(-\frac{\gamma}{4}\sum^{NT}_{m=1}\sum^N_{n=1}\lambda_{s_n}|\xi_{n,m}|^2\right).
\end{equation}
With comparison of (\ref{3.6}) and (\ref{3.7}), it is obvious to
note that the eigenvalue of the randomized matrix is the difference,
which suggests that employing a randomized matrix for a STC scheme
at the relay node can provide an improvement in BER performance.

\section{Simulations}

The simulation results are provided in this section to assess the
proposed scheme and algorithms. The cooperative MIMO system
considered employs an AF protocol with the Alamouti STBC scheme
\cite{Hamid} using QPSK modulation in a quasi-static block fading
channel with AWGN, as described in Section II. The bit error ratio
(BER) performance of the ARMO algorithm is assessed. The simulation
system with 1 relay node and each transmitting and receiving node
employs 2 antennas. In the simulations, we define both the symbol
power at the source node and the noise variance $\sigma^{2}$ for
each link as equal to 1.

The upper bounds of the D-Alamouti and the randomized D-Alamouti
derived in the previous section are shown in Fig. 2. The theoretical
pairwise error probabilities provide the largest decoding errors of
the two different coding schemes and as shown in the figure, by
employing a randomized matrix at the relay node it decreases the
decoding error upper bound. The comparison of the simulation results
in BER performance of the R-Alamouti and the D-Alamouti indicates
the advantage of using the randomized matrix.


The proposed ARMO algorithm is compared with the SM scheme and the
traditional RSTC algorithm using the distributed-Alamouti
(D-Alamouti) STBC scheme in \cite{RC De Lamare} with $n_r = 1$ relay
nodes in Fig. 3. The number of antennas $N=2$ at each node and the
effect of the direct link are considered. The results illustrate
that without the direct link, by making use of the STC or the RSTC
technique, a significant performance improvement can be achieved
compared to the spatial multiplexing system.  The RSTC algorithm
outperforms the STC-AF system, while the ARMO algorithm can improve
the performance by about 3dB as compared to the RSTC algorithm. With
the consideration of the direct link, the results indicate that the
cooperative diversity order can be increased, and using the ARMO
algorithm achieves an improved performance with $2$dB of gain as
compared to employing the RSTC algorithm and $3$dB of gain as
compared to employing the traditional STC-AF algorithm.


The simulation results shown in Fig. 4 illustrate the convergence
property of the ARMO algorithm. The SM, D-Alamouti and the
randomized D-Alamouti algorithms obtain nearly flat performance in
BER as the utilization of fixed STC scheme and the randomized
matrix. The SM scheme has the worst performance due to the lack of
coding gains, while the D-Alamouti scheme can provide a significant
performance improvement in terms of the BER improvement, and by
employing the randomized matrix at the relay node the BER
performance can decrease further when the transmission circumstances
are the same as that of the D-Alamouti. The ARMO algorithm shows its
advantage in a fast convergence and a lower BER achievement. At the
beginning of the optimization process with a small number of
samples, the ARMO algorithm achieves the BER level of the D-Alamouti
one, but with the increase of the received symbols, the ARMO
algorithm achieves a better BER performance.

\section{Conclusion}

We have proposed an adaptive randomized matrix optimization (ARMO)
algorithm for the randomized DSTC using a linear MMSE receive filter
at the destination node. The pairwise error probability of
introducing the randomized DSTC in a cooperative MIMO network with
the AF protocol has been derived. The simulation results illustrate
the advantage of the proposed ARMO algorithm by comparing it with
the cooperative network employing the traditional DSTC scheme and
the fixed randomized STC scheme. The proposed algorithm can be used
with different distributed STC schemes using the AF strategy and can
also be extended to the DF cooperation protocol.




\bibliographystyle{IEEEtran}

\end{document}